\documentclass[]{aastex631}

\usepackage{gensymb}

\graphicspath{{./}{figures/}}

\begin{document}

\title{Quick-Look Pipeline Light Curves for 5.7 Million Stars Observed Over the Second Year of TESS' First Extended Mission}

\correspondingauthor{Michelle Kunimoto}
\email{mkuni@mit.edu}

\author[0000-0001-9269-8060]{Michelle Kunimoto}
\affiliation{Department of Physics and Kavli Institute for Astrophysics and Space Research, Massachusetts Institute of Technology, Cambridge, MA 02139}

\author[0000-0002-5308-8603]{Evan Tey}
\affiliation{Department of Physics and Kavli Institute for Astrophysics and Space Research, Massachusetts Institute of Technology, Cambridge, MA 02139}

\author[0000-0003-0241-2757]{Willie Fong}
\affiliation{Department of Physics and Kavli Institute for Astrophysics and Space Research, Massachusetts Institute of Technology, Cambridge, MA 02139}

\author[0000-0002-2135-9018]{Katharine Hesse}
\affiliation{Department of Physics and Kavli Institute for Astrophysics and Space Research, Massachusetts Institute of Technology, Cambridge, MA 02139}

\author[0000-0002-1836-3120]{Avi Shporer}
\affiliation{Department of Physics and Kavli Institute for Astrophysics and Space Research, Massachusetts Institute of Technology, Cambridge, MA 02139}

\author[0000-0002-9113-7162]{Michael Fausnaugh}
\affiliation{Department of Physics and Kavli Institute for Astrophysics and Space Research, Massachusetts Institute of Technology, Cambridge, MA 02139}

\author[0000-0001-6763-6562]{Roland Vanderspek}
\affiliation{Department of Physics and Kavli Institute for Astrophysics and Space Research, Massachusetts Institute of Technology, Cambridge, MA 02139}

\author[0000-0003-2058-6662]{George Ricker}
\affiliation{Department of Physics and Kavli Institute for Astrophysics and Space Research, Massachusetts Institute of Technology, Cambridge, MA 02139}

\begin{abstract}
We present High-Level Science Products (HLSPs) containing light curves from MIT's Quick-Look Pipeline (QLP) from the second year of TESS' first Extended Mission (Sectors 40 -- 55; 2021 July -- 2022 September). In total, 12.2 million per-sector light curves for 5.7 million unique stars were extracted from 10-minute cadence Full-Frame Images (FFIs) and are made available to the community. As in previous deliveries, QLP HLSPs include both raw and detrended flux time series for all observed stars brighter than TESS magnitude $T = 13.5$ mag. Starting in Sector 41, QLP also produces light curves for select fainter M dwarfs. QLP has provided the community with one of the largest sources of FFI-extracted light curves to date since the start of the TESS mission.
\end{abstract}


\section{Introduction}

NASA's Transiting Exoplanet Survey Satellite \cite[TESS;][]{Ricker2015} launched in April 2018 with the goal of searching for transiting exoplanets around nearby, bright stars. Roughly $20,000$ pre-selected stars receive 2-minute cadence observations every 27.4-day sector, which are processed by the TESS Science Processing Operations Center \citep[SPOC;][]{Jenkins2016} at the NASA Ames Research Center. However, TESS also enables the discovery of planets around millions more stars by recording measurements of its entire $96\degree\times24\degree$ field of view in Full-Frame Images (FFIs). These FFIs were sampled every 30 minutes during the TESS Prime Mission (2018 July - 2020 July) and every 10 minutes during the first Extended Mission (2020 July - 2022 September). Relevant for exoplanet searches, the faster 10-minute cadence better resolves transit shapes and improves the detectability of short-duration signals.

In \citet{Kunimoto2021}, we presented light curves extracted from the FFIs by MIT's Quick-Look Pipeline \citep[QLP;][]{Huang2020} for 9.1 million stars observed in the first year of the first Extended Mission. These observations targeted the southern ecliptic hemisphere and covered Sectors 27 -- 39. Here, we present light curves from the second year of the first Extended Mission, which targeted the northern hemisphere and the ecliptic plane and covered Sectors 40 -- 55. In total, 12.2 million individual light curves for 5.7 million unique stars are available as High-Level Science Products (HLSPs) on the Mikulski Archive for Space Telescopes (MAST): \dataset[10.17909/t9-r086-e880]{http://dx.doi.org/10.17909/t9-r086-e880}.

\section{Updates}

The full QLP light curve extraction and post-processing procedure is described in \citet{Huang2020} and \citet{Kunimoto2021}. The procedure remains unchanged, with the exception of the following:

\begin{enumerate}
\item \textbf{Target Selection:} QLP produces light curves for all stars brighter than $T = 13.5$ mag as listed in the TESS Input Catalog \cite[currently TICv8.2;][]{Paegert2021}. Starting in Sector 41, the target selection was adjusted to include fainter M dwarfs ($T < 15$ mag, $T_{\rm eff} < 4000$ K, $R_{\star} < 0.8~R_{\odot}$), resulting in an additional $\sim$370,000 stars receiving light curves beyond the magnitude-limited sample. This decision had two primary motivations: (a) the small sizes of M dwarfs result in relatively large planet-to-star radius ratios, enabling the detection of planets around even very faint stars with poor photometric precision; and (b) stellar properties have improved since the earliest versions of the TIC used by TESS and QLP \citep{Stassun2018}, enabling more robust identification of M dwarfs.
\item \textbf{Use of TICA HLSPs:} The first step of QLP photometry is to calibrate the FFIs using the TESS Image CAlibrator \cite[TICA;][]{Fausnaugh2020}. Since the start of the TESS mission, this calibration was performed using a local installation of TICA (v0.2.1 for Sectors 40 -- 48). Starting in Sector 49, QLP uses the same TICA-calibrated FFIs as delivered to MAST (v1.0.2+ for Sectors 49 -- 55). The TICA version used is given in the \textbf{CALIB} header of our HLSPs. Small adjustments were made to the FFI linearity correction and smear correction between TICA v0.2.1 and v1.0.1.  To test the consequences of these changes, we compared the original QLP Sector 47 light curves for $\sim$23,000 stars with new light curves produced from the HLSPs. We measured the median difference in normalized flux values for each star. These differences were consistently below expected light curve precision, with typical values of $\sim$4 - 300 ppm depending on host star magnitude, and thus we do not consider their effects significant.
\item \textbf{Global Background Subtraction:} The second step of QLP photometry is to perform a global background subtraction for all calibrated images in order to remove significant scattered light. Global background subtraction was performed using \texttt{nebuliser} \citep{Irwin1985} since the start of the TESS mission. However, this occasionally produced discontinuities in QLP light curves, most notably in Sector 39. This step was temporarily disabled in Sector 40 and the first orbit of Sector 41, and re-enabled in the second orbit of Sector 41 using \texttt{fitsh} \citep{Pal2012} instead of \texttt{nebuliser}. Scattered light and small-scale features are still removed in the local background subtraction step later in the pipeline, which has remained unchanged, and thus disabling global background subtraction resulted in only minor reduction to light curve quality.

\end{enumerate}

\section{QLP High Level Science Products}

QLP HLSPs are FITS files that contain the raw and detrended light curves from aperture photometry based on an optimal aperture, and additional raw light curves from one smaller and one larger apertures. A full description of these FITS files is available in \citet{Huang2020} and \citet{Kunimoto2021}.

\begin{figure}
    \centering
    \includegraphics[width=0.49\textwidth]{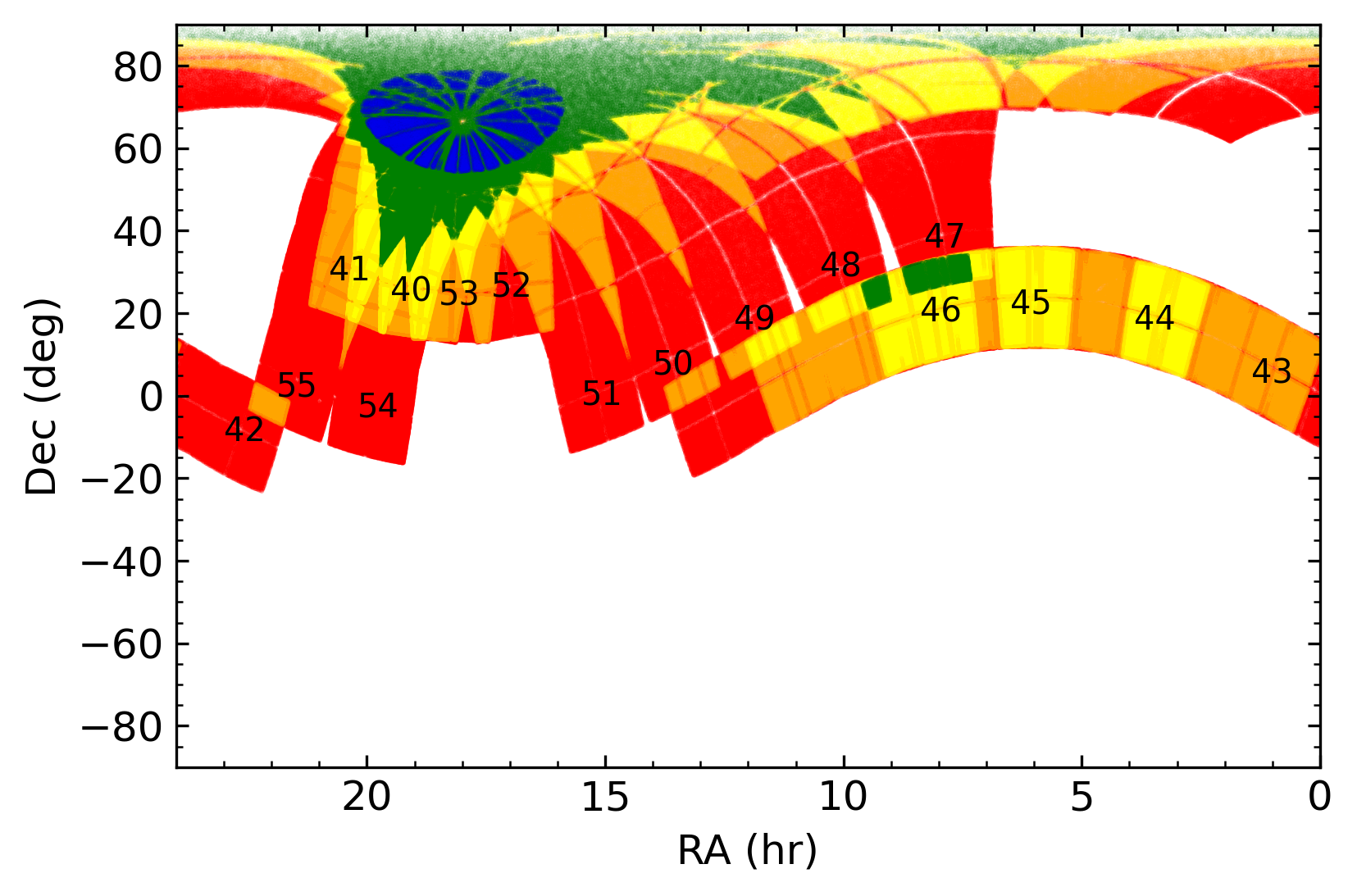}
    \includegraphics[width=0.49\textwidth]{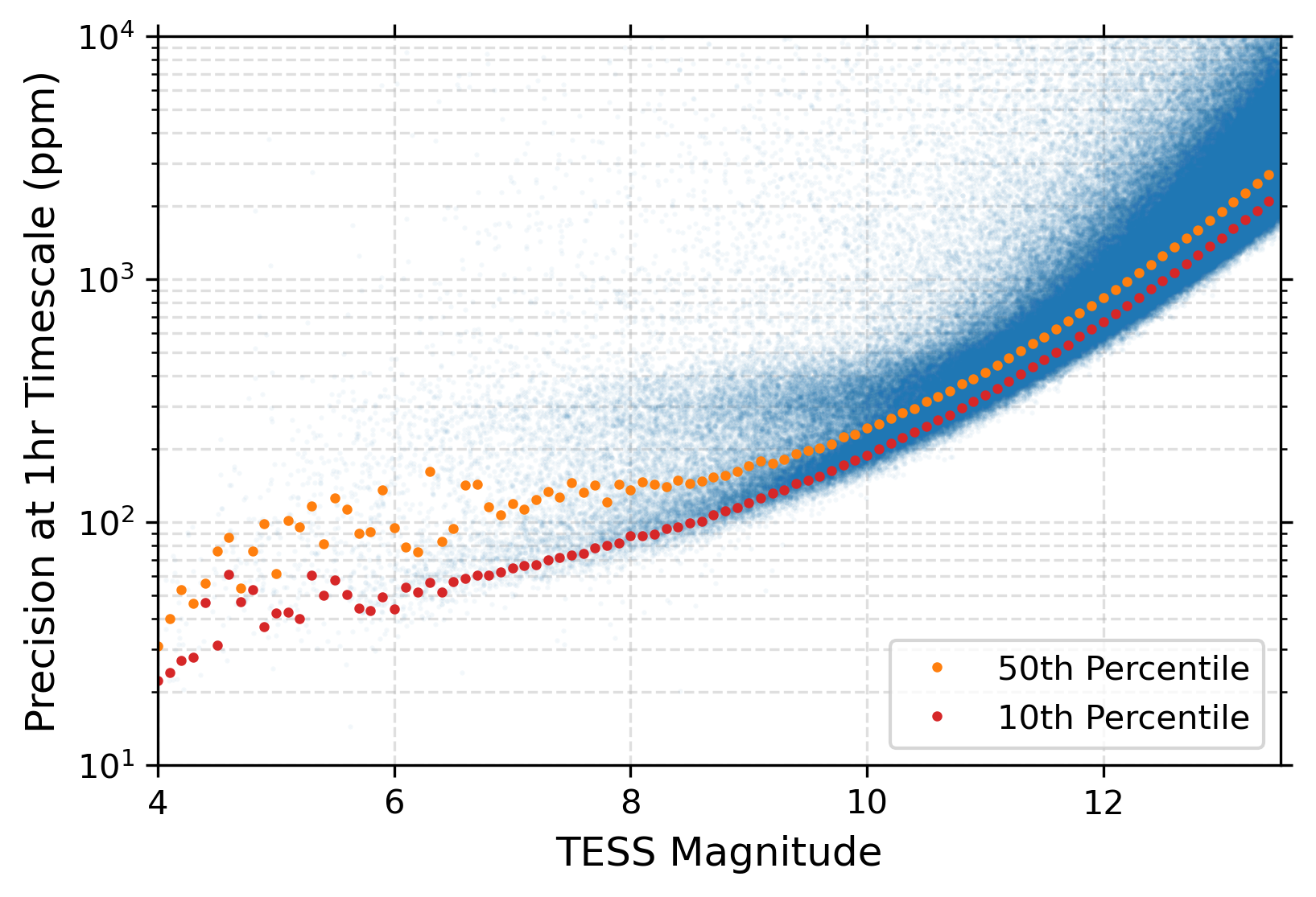}
    \caption{\textbf{Left:} The celestial coordinates of targets with QLP light curves from the northern ecliptic hemisphere (Sectors 40, 41, and 47 -- 55) and ecliptic (Sectors 42 -- 46), with sector numbers annotated over the centers of Camera 1. Targets are color-coded by the number of observed sectors, either 1 (red), 2 (orange), 3 (yellow), 4 -- 10 (green), or 11 (blue). \textbf{Right:} 1hr photometric precision of 16,000 randomly selected light curves from each orbit across Sectors 40 -- 55. The 50th and 10th percentiles of the precision binned every $\Delta T = 0.1$ mag are shown in orange and red, respectively.}
    \label{fig}
\end{figure}

Sky coverage across Sectors 40 -- 55 is shown in the left panel of Figure \ref{fig}. The right panel shows the precision of 16,000 randomly selected detrended light curves from each orbit as a function of TESS magnitude, based on light curves made with a 2.5-pixel circular aperture. The precision is estimated as 1.48 times the median absolute deviation of the light curve binned to a 1hr timescale.

\section{Acknowledgements}
This research note provides processing updates about the Quick Look Pipeline (QLP) operated by the TESS Science Office (TSO) at MIT. QLP extracts and detrends light curves from TESS Full-Frame Images (FFIs), searches light curves for transits, and produces vetting reports for promising planet candidates. The full QLP procedure is described in \citet{Huang2020, Kunimoto2021}. This work makes use of FFIs calibrated by TESS Image CAlibrator \citep[TICA;][]{Fausnaugh2020}, which are also available as High-Level Science Products (HLSPs) stored on the Mikulski Archive for Space Telescopes (MAST). Funding for the TESS mission is provided by NASA's Science Mission Directorate.

\bibliography{refs}

\end{document}